\documentclass[10pt]{article}
\usepackage{graphicx}
\usepackage{amsmath}
\usepackage{amssymb}
\usepackage{caption2}
\begin{document}
\bibliographystyle{prsty}
\begin{center}
{\large {\bf \sc{Light axial-vector tetraquark state  candidate: $a_1(1420)$ }}} \\[2mm]
Zhi-Gang  Wang \footnote{E-mail: zgwang@aliyun.com.  }     \\
 Department of Physics, North China Electric Power University, Baoding 071003, P. R. China \\
\end{center}

\begin{abstract}
In this article, we study  the   axial-vector  tetraquark state and two-quark-tetraquark mixed state consist of light quarks   using  the QCD sum rules. The present predictions  disfavor  assigning the $a_1(1420)$ as the axial-vector tetraquark state with $J^{PC}=1^{++}$, while support assigning the $a_1(1420)$ as the axial-vector two-quark-tetraquark mixed state.
\end{abstract}



\section{Introduction}

Recently  the  COMPASS collaboration   observed  the $a_1(1420)$ in the $f_0(980)\pi$ final state in the reaction $\pi^-+ p\to \pi^-\pi^-\pi^++p_{\rm recoil}$, the measured mass and width are 1420 MeV and 140 MeV, respectively \cite{COMPASS1312}. Although there are controversies about the quark structures of  the $f_0(980)$, it contains some $s\bar{s}$ constituents, irrespective of the two-quark state, molecular state or tetraquark state assignments \cite{Close2002}.   The $a_1(1420)$ is excellent candidate of the axial-vector tetraquark state with   the symmetric spin structure
$[su]_{S=1}[\bar{s}\bar{d}]_{S=0} + [su]_{S=0}[\bar{s}\bar{d}]_{S=1}$. On the other hand, the axial-vector meson $f_1(1420)$ with $J^{PC}=1^{++}$  has the mass and width $(1426.4\pm 0.9)\,\rm{MeV}$ and $(54.9\pm 2.6)\,\rm{MeV}$, respectively \cite{PDG}. The width of the $f_{1}(1420)$ is much smaller than that of the $a_1(1420)$,
it is unlikely that the $a_1(1420)$ and $f_1(1420)$ are the same particle.

 In this article, we take the $a_1(1420)$ as the axial-vector tetraquark state, study its mass and pole residues using the QCD sum rules,
 and make   possible assignment of the $a_1(1420)$ in the tetraquark scenario.
The QCD sum rules is a powerful theoretical tool is studying the ground state hadrons and has given many successful descriptions of the masses, decay constants, form-factors,   coupling constants, etc   \cite{SVZ79,Reinders85}.

The article is organized as:  we derive the QCD sum rules for
  the axial-vector tetraquark states  in section 2; in section 3 we present  numerical
results and discussions; while the last section  is reserved for our conclusion.

\section{QCD sum rules for  the  axial-vector tetraquark states }
We write down  the two-point correlation functions $\Pi_{\mu\nu}(p)$  in the QCD sum rules firstly,
\begin{eqnarray}
\Pi_{\mu\nu}(p)&=&i\int d^4x e^{ip \cdot x} \langle0|T\left\{J_\mu(x)J_\nu^{\dagger}(0)\right\}|0\rangle \, , \\
J^1_\mu(x)&=&\frac{\epsilon^{ijk}\epsilon^{imn}}{\sqrt{2}}\left\{u^j(x)C\gamma_5s^k(x) \bar{d}^m(x)\gamma_\mu C \bar{s}^n(x)+u^j(x)C\gamma_\mu s^k(x)\bar{d}^m(x)\gamma_5C \bar{s}^n(x) \right\} \, , \nonumber\\
J^2_\mu(x)&=&\frac{\epsilon^{ijk}\epsilon^{imn}}{\sqrt{2}}\left\{u^j(x)C\gamma_5s^k(x) \bar{d}^m(x)\gamma_\mu C \bar{s}^n(x)-u^j(x)C\gamma_\mu s^k(x)\bar{d}^m(x)\gamma_5C \bar{s}^n(x) \right\} \, , \nonumber\\
J^3_\mu(x)&=&cos\theta\,J^1_{\mu}(x)+sin\theta \, \frac{2\sqrt{2}}{3} \langle\bar{s}s \rangle \bar{d}(x)\gamma_5\gamma_\mu u(x) \, ,
\end{eqnarray}
where the currents  $J_\mu(x)=J^1_\mu(x),\, J^2_\mu(x),\,J^3_\mu(x)$,  the $i$, $j$, $k$, $m$, $n$ are color indexes, the $C$ is the charge conjunction matrix. The factor $\frac{2\sqrt{2}}{3} \langle\bar{s}s \rangle$ is added to normalize  the current $J^3_\mu(x)$, the $\theta$ is the mixing angle.
The currents $J^1_\mu(x)$ and $J^3_\mu(x)$ have the quantum numbers  $J^{PC}=1^{++}$, we choose them to interpolate
    the tetraquark state  and two-quark-tetraquark mixed state, respectively. The current $J^2_\mu(x)$ has the quantum numbers  $J^{PC}=1^{+-}$, we choose it to study
      the negative charge conjunction partner of the axial-vector tetraquark state as a byproduct.

At the hadronic representation, the ground states contributions from the axial-vector mesons can be written as
\begin{eqnarray}
\Pi_{\mu\nu}(p)&=&\frac{\lambda^2}{M^2-p^2}\left(-g_{\mu\nu} +\frac{p_\mu p_\nu}{p^2}\right) +\cdots \, \, ,
\end{eqnarray}
where the pole residues  $\lambda$ are defined by
\begin{eqnarray}
 \langle 0|J_\mu(0)|A(p)\rangle=\lambda \, \varepsilon_\mu \, ,
\end{eqnarray}
the $\varepsilon_\mu$ are the polarization vectors of the axial-vector mesons $A$.

We calculate  the contributions of the vacuum condensates up to dimension-10 in the operator product expansion, assume  vacuum saturation to factorize the  higher dimension vacuum condensates to lower dimension vacuum condensates, and obtain the spectral densities at the level of quark-gluon degree's of freedom  through  dispersion relation.  Then we take the
quark-hadron duality below the continuum thresholds $s_0$ and perform Borel transform  with respect to
the variable $-p^2$ to obtain three QCD sum rules,
\begin{eqnarray}
\lambda^2\, e^{-\frac{M^2}{T^2}}= \int_{0}^{s_0} dt\, \rho_i(t) \, e^{-\frac{t}{T^2}} \, ,
\end{eqnarray}
where the $\rho_i(t)$ with  $i=1,2,3$ are the QCD spectral densities corresponding to the interpolating currents $J^1_\mu(x)$, $J^2_\mu(x)$, $J^3_\mu(x)$, respectively,
\begin{eqnarray}
\rho_1(t)&=& \frac{t^4}{73728\pi^6}-\frac{7m_s\langle\bar{q}q\rangle t^2}{768\pi^4}+\frac{m_s\langle\bar{s}s\rangle t^2}{256\pi^4}+\frac{5m_s\langle\bar{q}g_s\sigma Gq\rangle t}{384\pi^4}-\frac{m_s\langle\bar{s}g_s\sigma Gs\rangle t}{288\pi^4}+\frac{ 5\langle\bar{q} q\rangle\langle\bar{s} s\rangle t}{72\pi^2} \nonumber\\
&&-\frac{ \langle\bar{q} q\rangle\langle\bar{s}g_s\sigma G s\rangle+ \langle\bar{s} s\rangle\langle\bar{q}g_s\sigma G q\rangle }{32\pi^2} +\frac{m_s\langle\bar{q} q\rangle\langle\bar{s} s\rangle^2}{18}\delta(t)-\frac{2m_s\langle\bar{q} q\rangle^2\langle\bar{s} s\rangle}{9}\delta(t)\nonumber\\
&&+\frac{ \langle\bar{q}g_s\sigma G q\rangle\langle\bar{s}g_s\sigma G s\rangle }{192\pi^2}\delta(t)+\frac{t^2}{9216\pi^4}\langle\frac{\alpha_sGG}{\pi}\rangle+\frac{\langle\bar{q} q\rangle\langle\bar{s} s\rangle}{144}\langle\frac{\alpha_sGG}{\pi}\rangle\delta(t)\nonumber\\
&&-\frac{\left[\langle\bar{q} q\rangle-\langle\bar{s} s\rangle\right]^2}{1296}\langle\frac{\alpha_sGG}{\pi}\rangle\delta(t)+\frac{5m_s\left[\langle\bar{q}g_s\sigma G q\rangle-\langle\bar{s}g_s\sigma G s\rangle \right]t}{3456\pi^4} \nonumber\\
&&+\frac{\left[\langle\bar{q} q\rangle-\langle\bar{s} s\rangle \right]\left[\langle\bar{q}g_s\sigma G q\rangle-\langle\bar{s}g_s\sigma G s\rangle \right]}{288\pi^2}
-\frac{\left[\langle\bar{q}g_s\sigma G q\rangle-\langle\bar{s}g_s\sigma G s\rangle \right]^2}{1728\pi^2}\delta(t) \, ,
\end{eqnarray}

\begin{eqnarray}
\rho_2(t)&=& \frac{t^4}{73728\pi^6}-\frac{7m_s\langle\bar{q}q\rangle t^2}{768\pi^4}+\frac{m_s\langle\bar{s}s\rangle t^2}{256\pi^4}+\frac{5m_s\langle\bar{q}g_s\sigma Gq\rangle t}{384\pi^4}-\frac{m_s\langle\bar{s}g_s\sigma Gs\rangle t}{288\pi^4}+\frac{ 5\langle\bar{q} q\rangle\langle\bar{s} s\rangle t}{72\pi^2} \nonumber\\
&&-\frac{ \langle\bar{q} q\rangle\langle\bar{s}g_s\sigma G s\rangle+ \langle\bar{s} s\rangle\langle\bar{q}g_s\sigma G q\rangle }{32\pi^2} +\frac{m_s\langle\bar{q} q\rangle\langle\bar{s} s\rangle^2}{18}\delta(t)-\frac{2m_s\langle\bar{q} q\rangle^2\langle\bar{s} s\rangle}{9}\delta(t)\nonumber\\
&&+\frac{ \langle\bar{q}g_s\sigma G q\rangle\langle\bar{s}g_s\sigma G s\rangle }{192\pi^2}\delta(t)+\frac{t^2}{9216\pi^4}\langle\frac{\alpha_sGG}{\pi}\rangle+\frac{\langle\bar{q} q\rangle\langle\bar{s} s\rangle}{144}\langle\frac{\alpha_sGG}{\pi}\rangle\delta(t)\nonumber\\
&&+\frac{\left[\langle\bar{q} q\rangle-\langle\bar{s} s\rangle\right]^2}{1296}\langle\frac{\alpha_sGG}{\pi}\rangle\delta(t)-\frac{5m_s\left[\langle\bar{q}g_s\sigma G q\rangle-\langle\bar{s}g_s\sigma G s\rangle \right]t}{3456\pi^4} \nonumber\\
&&-\frac{\left[\langle\bar{q} q\rangle-\langle\bar{s} s\rangle \right]\left[\langle\bar{q}g_s\sigma G q\rangle-\langle\bar{s}g_s\sigma G s\rangle \right]}{288\pi^2}
+\frac{\left[\langle\bar{q}g_s\sigma G q\rangle-\langle\bar{s}g_s\sigma G s\rangle \right]^2}{1728\pi^2}\delta(t) \, ,
\end{eqnarray}

\begin{eqnarray}
\rho_3(t)&=&cos^2\theta\, \rho_1(t)+sin^2\theta\, \frac{8\langle\bar{s}s\rangle^2}{9}\left\{ \frac{t^2}{4\pi^2}-\frac{1}{12}\langle\frac{\alpha_sGG}{\pi}\rangle\delta(t) \right\} \, .
\end{eqnarray}
The contribution of the three gluon condensate $\langle g_s^3 GGG\rangle$ is of the order $\mathcal{O}(\sqrt{\alpha_s}^{3})$ and   numerically very small, so it is neglected in this article.
The differences  between the spectral densities $\rho_1(t)$ and $\rho_2(t)$ are proportional to $\left[\langle\bar{q}g_s\sigma G q\rangle-\langle\bar{s}g_s\sigma G s\rangle \right]$, $\left[\langle\bar{q} q\rangle-\langle\bar{s} s\rangle\right]^2$, $\left[\langle\bar{q} q\rangle-\langle\bar{s} s\rangle \right]\left[\langle\bar{q}g_s\sigma G q\rangle-\langle\bar{s}g_s\sigma G s\rangle \right]$, $\left[\langle\bar{q}g_s\sigma G q\rangle-\langle\bar{s}g_s\sigma G s\rangle \right]^2$, which are of minor importance and lead to almost  degenerate masses for the charge conjunction partners.

We can differentiate  Eq.(5) with respect to  $\frac{1}{T^2}$, then eliminate the
 pole residues $\lambda$ so as to  obtain the QCD sum rules for
 the masses of the axial-vector mesons,
 \begin{eqnarray}
 M^2= \frac{\int_{0}^{s_0} dt\frac{d}{d \left(-1/T^2\right)}\rho_i(t)e^{-\frac{t}{T^2}}}{\int_{0}^{s_0} dt  \rho_i(t)e^{-\frac{t}{T^2}}}\, .
\end{eqnarray}

\section{Numerical results and discussions}
The basic input parameters at the operator product expansion  side  are taken as $m_s=0.118\,\rm{GeV}$, $\langle
\bar{q}q \rangle=-(0.24\pm 0.01\, \rm{GeV})^3$,  $\langle \bar{s}s \rangle=(0.8\pm0.1)\langle \bar{q}q \rangle$, $\langle
\bar{q}g_s\sigma G q \rangle=m_0^2\langle \bar{q}q \rangle$, $\langle
\bar{s}g_s\sigma G s \rangle=m_0^2\langle \bar{s}s \rangle$,
$m_0^2=(0.8 \pm 0.1)\,\rm{GeV}^2$, $\langle \frac{\alpha_s
GG}{\pi}\rangle=(0.33\,\rm{GeV})^4 $     at the energy scale  $\mu=1\, \rm{GeV}$
\cite{PDG,SVZ79,Reinders85,Ioffe2005,ColangeloReview}.
We usually take the following  two
criteria of the QCD sum rules:\\
 $\bullet$ Pole dominance at the phenomenological side;\\
 $\bullet$ Convergence of the operator product expansion \cite{Wang-NPA}, \\
 to choose the Borel parameters and continuum threshold parameters. It is difficult to satisfy the two criteria as the conventional two-quark mesons for the light tetraquark states \cite{Wang-NPA}, we have to release either of the two criteria. In other words, we can retain pole dominance by choosing small Borel parameters or retain convergence of the operator product expansion by choosing large Borel parameters.

 In calculations, we observe that there appear Borel platforms (or minimum) at a special value below $T^2=1.0\,\rm{GeV}^2$, then the hadronic masses increase  monotonously with increase of the Borel parameters, although  the curves of the line-shapes  are not steep in some cases. On the other hand, the pole dominance is well satisfied at the region $T^2\leq 1\,\rm{GeV}^2$.

 If we choose the Borel parameters at the region $T^2\leq 1\,\rm{GeV}^2$, the dominant contributions come from the  $\langle\bar{q}{q}\rangle\langle\bar{s}{s}\rangle$ and $\langle\bar{q}{q}\rangle\langle\bar{s}g_s \sigma G{s}\rangle+\langle\bar{s}{s}\rangle\langle\bar{q}g_s \sigma G{q}\rangle$ terms rather than  the perturbative terms, the vacuum condenses ($\langle\bar{q}{q}\rangle\langle\bar{s}{s}\rangle \langle \frac{\alpha_sGG}{\pi}\rangle$, $\langle\bar{q}g_s \sigma G{q}\rangle\langle\bar{s}g_s \sigma G{s}\rangle$, etc) of dimension 10 are of minor importance.

 In this article, we retain the criterium of pole dominance and  modify the criterium of convergence of the operator product expansion to be the contributions of the vacuum condensates of dimension 10 are less (or much less) than $15\%$ of that of the $\langle\bar{q}{q}\rangle\langle\bar{s}{s}\rangle$. Then the operator product expansion is still convergent, but the convergent behavior is slower than that of the conventional two-quark mesons.

\begin{table}
\begin{center}
\begin{tabular}{|c|c|c|c|c|c|c|c|}\hline\hline
$J^{PC}$                   &$T^2 (\rm{GeV}^2)$ &$\sqrt{s_0} (\rm{GeV})$ &pole         & $M(\rm{GeV})$           & $\lambda(\rm{GeV}^5)$ \\ \hline
$1^{++}$                   &$0.8-1.1$          &$2.3\pm0.1$             &$(66-92)\%$  & $1.83^{+0.15}_{-0.10}$  & $4.29^{+1.34}_{-0.60}\times10^{-3}$ \\ \hline
$1^{+-}$                   &$0.8-1.1$          &$2.3\pm0.1$             &$(66-92)\%$  & $1.82^{+0.15}_{-0.09}$  & $4.27^{+1.24}_{-0.60}\times10^{-3}$ \\ \hline
$1^{++}(\theta=25^\circ)$  &$0.5-0.7$          &$1.9\pm0.1$             &$(81-95)\%$  & $1.42^{+0.15}_{-0.10}$  & $2.35^{+0.71}_{-0.34}\times10^{-3}$ \\ \hline
 \hline
\end{tabular}
\end{center}
\caption{ The Borel parameters, continuum threshold parameters, pole contributions, masses and pole residues of the axial-vector mesons. }
\end{table}

 The resulting Borel parameters, continuum threshold parameters and the pole contributions are shown explicitly in Table 1. We take into account all uncertainties of the input parameters,
and obtain the  masses and pole residues of
 the   axial-vector mesons, which are  shown Table 1 and  Fig.1.

 From Table 1, we can see that the tetraquark states with the $J^{PC}=1^{++}$ and $1^{+-}$ have almost degenerate masses,  the values $1.83^{+0.15}_{-0.10}\,\rm{GeV}$ and $1.82^{+0.15}_{-0.09}\,\rm{GeV}$ are above the experimental value $M_{a_1(1420)}=1.42\,\rm{GeV}$. The numerical results  disfavor  assigning the $a_1(1420)$ as the axial-vector tetraquark state. The mass of the axial-vector meson $a_1(1260)$ is $M_{a_1(1260)}=(1230\pm40)\,\rm{MeV}$ from the Particle Data Group \cite{PDG}. If we take the $a_1(1420)$ as the mixed state of the $a_1(1260)$ and tetraquark state $[su]_{S=1}[\bar{s}\bar{d}]_{S=0} + [su]_{S=0}[\bar{s}\bar{d}]_{S=1}$, the component $a_1(1260)$ can reduce the mass to the experimental value. In calculations, we observe that the mixing angle $\theta=25^{\circ}$ lead to the value $1.42^{+0.15}_{-0.10}\,\rm{GeV}$, which reproduces the experimental data.  The $a_1(1420)$ is produced through its $a_1(1260)$ component, then decays through its tetraquark component, $\pi^-+ p\to a_1(1420)+p_{\rm recoil} \to f_0(980)+\pi^{-}+p_{\rm recoil}\to \pi^-\pi^-\pi^++p_{\rm recoil}$.

\begin{figure}
\centering
\includegraphics[totalheight=6cm,width=7cm]{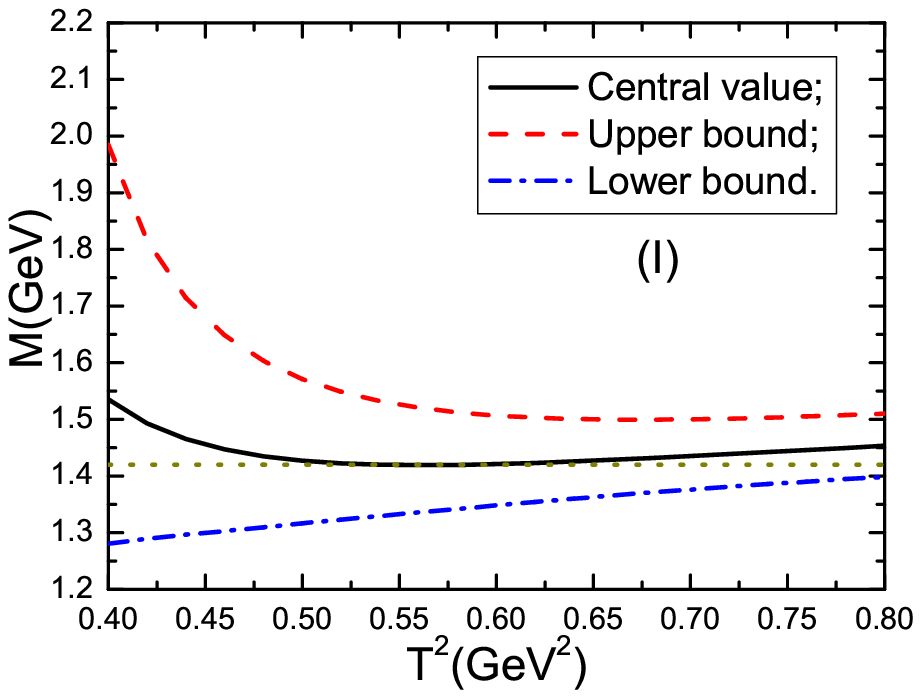}
\includegraphics[totalheight=6cm,width=7cm]{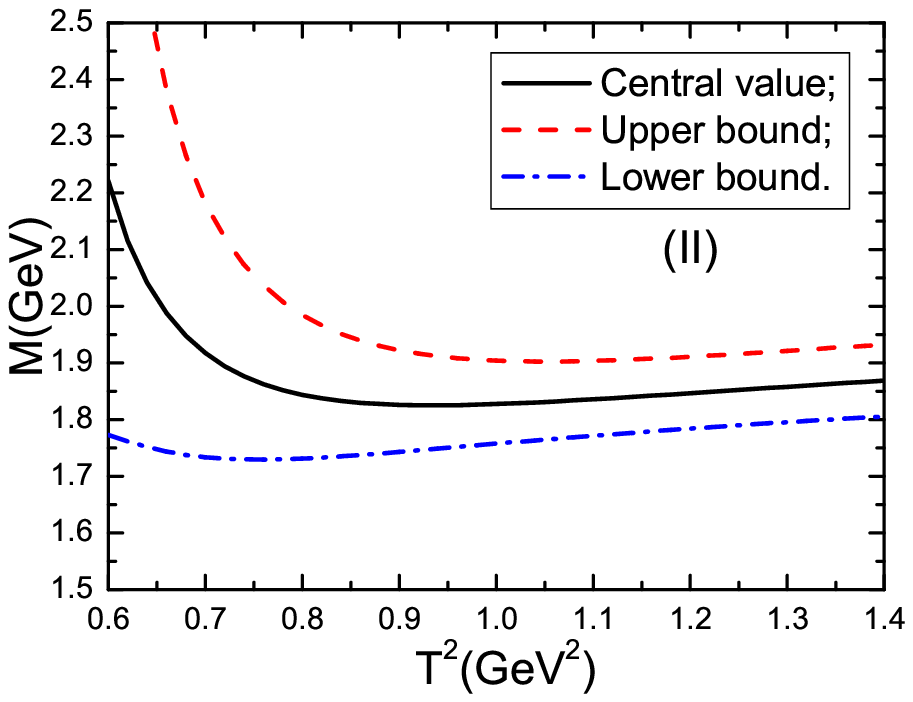}
  \caption{ The masses of the axial-vector mesons with variations of the  Borel parameters $T^2$, where the horizontal line denotes  the experimental value,
  the   (I) and (II) denote the  two-quark-tetraquark mixed state and tetraquark state with $J^{PC}=1^{++}$, respectively.}
\end{figure}

\section{Conclusion}
In this article, we study  the  axial-vector  tetraquark state and two-quark-tetraquark mixed state consist of light quarks  using  the QCD sum rules. The present predictions  disfavor  assigning the $a_1(1420)$ as the axial-vector tetraquark state with $J^{PC}=1^{++}$, while support assigning the $a_1(1420)$ as the axial-vector two-quark-tetraquark mixed state. Furthermore, we obtain the mass of the axial-vector tetraquark state with $J^{PC}=1^{+-}$ as a byproduct, which can be confronted with the experimental data in the future.

\section*{Acknowledgements}
This  work is supported by National Natural Science Foundation,
Grant Number 11375063,  and the Fundamental Research Funds for the
Central Universities.


\begin{thebibliography}{99}

\bibitem{COMPASS1312} S. Paul, arXiv:1312.3678.

\bibitem{Close2002}  C. Amsler and N. A. Tornqvist, Phys. Rept. {\bf 389} (2004) 61.

\bibitem{PDG}   J. Beringer et al, Phys. Rev. {\bf D86} (2012) 010001.

\bibitem{SVZ79}  M. A. Shifman, A. I. Vainshtein and V. I. Zakharov, Nucl. Phys. {\bf B147} (1979) 385.

\bibitem{Reinders85} L. J. Reinders, H. Rubinstein and S. Yazaki, Phys. Rept. {\bf 127} (1985) 1.



\bibitem{Wang-NPA} Z. G. Wang, Nucl. Phys. {\bf A791} (2007) 106.


\bibitem{Ioffe2005} B. L. Ioffe, Prog. Part. Nucl. Phys. {\bf 56} (2006) 232.

\bibitem{ColangeloReview} P. Colangelo and A. Khodjamirian, hep-ph/0010175.






\end{thebibliography}
\end{document}